\documentclass[11pt]{article}
\usepackage{graphicx}
\usepackage{amsmath}
\usepackage{amssymb}
\usepackage{amsxtra}
\usepackage[autostyle]{csquotes}
\usepackage{epstopdf}
\usepackage{float}

\title{Primordial Black Hole Clusters and their Evolution}

\author{M.Yu. Khlopov\\Centre for Cosmoparticle Physics "Cosmion"  \\
National Research Nuclear University\\ "Moscow Engineering Physics Institute", 115409 Moscow, Russia \\
APC laboratory 10, rue Alice Domon et L\'eonie Duquet \\75205
Paris Cedex 13, France\\
khlopov@apc.univ-paris7.fr\\N.A. Chasnikov\\
National Research Nuclear University\\ "Moscow Engineering Physics Institute", 115409 Moscow, Russia \\
nchasnikov@gmail.com}

\begin{document}
\maketitle

\begin{abstract}
A possibility of pregalactic seeds of the Active Galactic Nuclei can be a nontrivial cosmological consequence of particle theory.
Such seeds can appear as Primordial Black Hole (PBH) clusters, formed in the succession of phase transitions with spontaneous and then manifest breaking of the global U(1) symmetry. If the first phase transition takes place at the inflationary stage, a set of massive closed walls may be formed at the second phase transition and the collapse of these closed walls can result in formation of PBH clusters.
We present the results of our studies of the evolution of such PBH Clusters.
\end{abstract}
\section{Introduction}\label{intr}
Primordial Black Holes (PBHs) are a very sensitive cosmological
probe for physics phenomena occurring in the early Universe. They
could be formed by many different mechanisms, reflecting the fundamental structure of particle theory and nonhomogeneity of very early Universe. Here after a brief review of mechanisms of PBH formation we consider a nontrivial possibility of clusters of massive PBHs. The evolution of such clusters can provide pregalactic seeds of Active Galactic Nuclei (AGN) and we discuss various aspects of this evolution in the present paper.

\section{PBH Formation}\label{s:pbhf}
Primordial Black Holes can be formed in many different ways~\cite{pbh1,DMRev}, such as initial density inhomogeneities, first order and non-equilibrium second order phase transitions, etc. Let us give a brief review of these possibilities.

\subsection{PHB formation in initial density inhomogeneities}\label{s:pbhfidi}
A probability for fluctuation of ~ 1 for metric fluctuations distributed according to Gaussian law with dispersion $\left \langle \delta^{2} \right \rangle \ll 1$ is determined by exponentially small tail of high amplitude part of this distribution.
In non-Gaussian fluctuations, this process can be more suppressed~\cite{art11}.
In the space, described by
\begin{equation}\label{eq1}
p = \gamma \epsilon , 0\leq \gamma \leq 1
\end{equation}
equation of state a probability to form black hole from fluctuation within cosmological horizon is given by~\cite{bcph,newBook}
\begin{equation}\label{eq2}
W_{PBH}=e^{-\frac{\gamma^{2}}{2\left \langle \delta^{2} \right \rangle}}
\end{equation}
It provides exponential sensitivity of PBH spectrum to softening of equation of state in early Universe ($\gamma \rightarrow 0 $) or to increase of ultraviolet part of spectrum of density fluctuations ($\delta^{2} \rightarrow 1$).
These phenomena can appear as cosmological consequence of particle theory (see ~\cite{bcph,newBook} for review of this and some other mechanisms of PBH formation and for references).
\subsection{PBH from non-equilibrium second order phase transition}\label{s:pbhfidi2}
The mechanism of PBH formation in the non-equilibrium second order phase transition is of special interest, since it can provide formation of massive and even Supermassive PBHs. PBHs are produced in this mechanism by self-collapsing of closed domain walls. If there are two vacuum states of a system, there are two possibilities to populate that states in the early Universe: under the usual circumstances of thermal phase transition the Universe contains both states populated with equal probability. The other possibility is beyond the pure thermodynamical equilibrium condition, when the two vacuum states are populated with islands of the less probable vacuum, surrounded by the sea of another, more preferable, vacuum.

It is necessary to redefine effectively the correlation length of the scalar field that drives a phase transition and consequently the formation of topological defects and  the only necessary ingredient for that is the existence of an effectively flat direction(s), along which the scalar potential vanishes during inflation. 

The background deSitter fluctuations of such effectively massless scalar field could provide non-equilibrium redefinition of correlation length and give rise to the islands of one vacuum in the sea of another one. In spite of such redefinition the phase transition itself takes place deeply in the Friedman-Robertson-Walker (FRW) epoch. After the phase transition two vacua are separated by a wall, and such a closed wall, separating the island with the less probable vacuum, can be very large. 

At some moment after crossing horizon the walls start shrinking due to surface tension. As a result, if the wall does not release the significant fraction of its energy in the form of outward scalar waves, almost the whole energy of such closed wall can be concentrated in a small volume within its gravitational radius what is the necessary condition for PBH formation. 

The mass spectrum of the PBHs which can be created by such a way depends on the scalar field potential which parameterizes the flat direction during inflation and triggers the phase transition at
the FRW stage.

We consider the Universe that, due to the existence of an inflaton, goes through a period of inflation
and then settles down to the standard FRW geometry. Then we introduce a complex scalar
field $\varphi$, not the inflaton, with a large radial mass $\sqrt{\lambda f}> H_{i}$ that has got Mexican hat potential
\begin{equation}
V(\varphi )=\lambda\left( \left | \varphi  \right |^{2}-\frac{f^{2}}{2}  \right),
\end{equation}
which provides the U(1) symmetry spontaneous breaking in the period of inflation, corresponding to the scales of
the modern cosmological horizon. Therefore we deal only with the phase of that complex field $\theta = \frac{\varphi}{f}$, which parameterizes potential
\begin{equation}
V=\Lambda^{4}\left ( 1- cos\frac{\phi }{f} \right )
\end{equation}
Under this condition we come to the conclusion, that the correlation
length of second order phase transition with spontaneously broken U(1) symmetry exceeds the present
cosmological horizon, and all global U(1) strings are beyond our horizon. If we assume $m \ll H_{i}$ then this implies that during inflation the potential energy of field $\varphi$ is much smaller than the cosmological friction term what justifies neglecting the potential until the Universe goes deeply into the FRW phase.  During inflation and long afterward, $H_{i}$ is very large (by assumption) compared to the potential (2). It follows that we can drop the gradient term in the equation of motion~\cite{pbhso}
\begin{equation}
\ddot{\theta} + 3H\dot{\theta}+\frac{dV}{d\theta} = 0
\end{equation}
and resulting equation is solved by $\theta_{0} = \theta_{Nmax}$ , where $\theta_{Nmax}$ is an arbitrary constant. In the standard assumption, our present horizon has been nucleated at the $N_{max}$ e-folds before the end of inflationary epoch, being embedded in an enormous inflation horizon, created by exponential blow up of a single casual horizon.  It follows that $\theta_{Nmax}$ will be the same over the inter inflationary horizon. Without loss of generality, we put $\theta_{Nmax} < \pi$ and considering the quantum fluctuations of the phase $\theta$ at the deSitter background. There are quantum fluctuations produced on the vacuum state of $\theta$ due to the boundary conditions of deSitter space.
These fluctuations are sometimes referred to as contribution to the ``Hawking temperature'' of deSitter space but, there are no true thermal effects. It makes the dynamics of phase $\theta$ strongly non-equilibrium leading to the non-thermal distribution of scales populated with different vacuums in the postinflationary Universe.  The average amplitude of such fluctuations for massless field generated during each time interval $H_{_i}^{-1}$ is $\delta \theta =\frac{H_{i}}{2\pi f}$.  The total number of steps during time interval $\Delta t$ is given by $N = H_{i}\Delta t$ - looks like the one-dimensional Brownian motion.  Each domain is characterized by average phase value $\theta _{Nmax-1}=\theta _{Nmax}\pm \delta \theta$. 

In the half of these domains the phases evolve toward $\pi$ while in the other domains they move toward zero. This process is duplicated in each volume of size $H^{-1}$ during next e-fold. Now at any given
scale $l=k^{-1}$ the size of distribution of the phase value $\theta$ can be described by Gaussian law~\cite{pbhso3}
\begin{equation}
P(\theta_{l},l)=\frac{1}{\sqrt{2\pi}\sigma _{l}}exp\left (-\frac{(\theta_{Nmax}-\theta_{l})^{2}}{2\sigma _{l}^{2}}  \right )
\end{equation}
It is recommended for more information to address the papers~\cite{pbh1,pbhso,pbhso1,cl1}.

\subsection{Initial PBH Mass spectrum}\label{s:initmass}
Initial mass spectrum $n\left ( m,t=0 \right )$ depends of parameter $f$ and $\Lambda$. ~\cite{pbhso} In addition, is a numerical solution.
There is another way: one can describe this system by Ito's equation. One-dimensional Brownian motion in the terms of stochastic equations is an Ornstein-Uhlenbeck process.~\cite{oup1,oup2} Using this mathematical framework,
one can find the analytical solution of Ito's equation and obtain initial mass spectrum as an analytical formula $n_{0}=n_{f,\Lambda} \left ( m,t=0\right )$. This method is in the process of development.

\section{Clusters of PHBs}\label{s:pbhclusters}
According to the $2^{nd}$ order phase transitions mechanism, PBH appears as a sufficiently large cluster, which could collapse into one large Super Massive Black Hole (SMBH) - the Active Galactic Nucleus (AGN) of the future galaxy.
\subsection{PBH Cluster dynamics}\label{pbhclusterd}
By analogy with the star cluster ~\cite{scd1} with the difference that the black holes can merge into one, the following processes are significant:
\begin{itemize}
	\item BH collisions $\rightarrow BH $ merging and as a result $N_{BH}\rightarrow 1$
	\item Flying-out BH from cluster $\rightarrow$ reducing the mass of the cluster~\cite{foc1,foc2}
	\item Dynamical friction $\rightarrow$ lower Maxwell distribution~\cite{dynfr}
\end{itemize}

Dynamical friction mostly contributes into the $\left \langle \sigma \nu  \right \rangle$ of the collision process.~\cite{dynfr}
The equation describing the dynamics of the BH Cluster is a modification of Smoluchowski (or Kolmogorov-Feller) equation and runs as follows
\begin{equation}\label{eq3}
\begin{split}
n{}'=\int_{0}^{M}n\left (m-\mu ,t \right )\left \langle \sigma \nu  \right \rangle_{m-\mu,m}d\mu \\- n\left ( m,t \right )\left (  \int_{0}^{\infty} n\left ( \mu,t  \right ) \left \langle \sigma \nu  \right \rangle_{\mu,m} d\mu + \int_{0}^{\infty} n \left (\mu,t \right )\Lambda\left ( m,\mu \right ) \mu^{2} d\mu \right ),
\end{split}
\end{equation}
where M is total initial cluster mass, $\Lambda \left (m,\mu \right )$ one can find in ~\cite{foc2}.
Let us consider numerical solution of that equation for arbitrary initial parameters:
\includegraphics[width=12 cm,height=5.7cm]{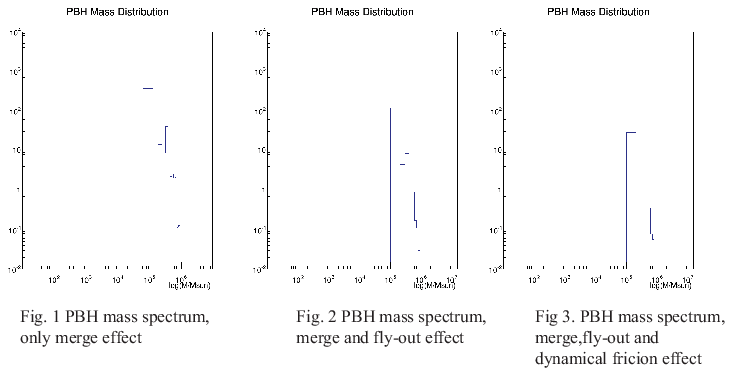}
\\Numerical solution of that equation shows, that there is the ``trend" to the decrease of BHs with larger masses.
However, the numerical solution is indispensable because of the unknown initial parameters $f$ and $\Lambda$ of the model.
 Exact analytical solution of that equation is overly precise and is a very nontrivial exercise.
 If a single SMBH is supposed to be the result, one needs to get a solution of that equation as
\begin{equation}
n(m,t)=\delta(m-m_{SMBH} )\chi(t-t_{gen} )
\end{equation}
Substituting this partial solution into the equation of the PBH cluster dynamics, one can obtain the timescale of the process and the mass of the resulting SMBH as a functionals of initial conditions:
\begin{equation}\label{eq4}
m_{SMBH}=F \left [ n \left(m,t=0\right) \right]
\end{equation}
\begin{equation}\label{eq5}
t_{gen}=G \left [n(m,t=0)\right ]
\end{equation}
The calculated values of $ t_{gen}$  and $m_{SMBH}$ can be confronted with the observational data, putting constraints on the fundamental physical scales $f$ and $\Lambda$.
\section*{Acknowledgements}
We are grateful to the Organizers of XVI Bled Workshop for hospitality and fruitful and creative atmosphere of discussions.


\end{document}